\documentclass[12pt]{article}
\usepackage{hepth}
\usepackage{graphicx}

\begin{document}

\preprint{PUPT 2286 \\ LMU-ASC 53-08}

\institution{LMU}{Ludwig-Maximilians-Universit\"at, Department f\"ur Physik, Theresienstrasse 37, \cr 80333 M\"unchen, Germany}

\institution{PU}{Joseph Henry Laboratories, Princeton University, Princeton, NJ 08544, USA}

\title{Universality of second order transport coefficients from the gauge-string duality}

\authors{Michael Haack\worksat{\LMU,}\footnote{e-mail: {\tt michael.haack@physik.uni-muenchen.de}} and
Amos Yarom\worksat{\PU,}\footnote{e-mail: {\tt ayarom@princeton.edu}}}

\abstract{We consider the strongly coupled limit of conformal gauge theory plasmas with conserved $U(1)$ charges which have a gravity dual. We show that, under mild restrictions, the second order transport coefficients of such theories satisfy a universal relation among themselves, similar to the shear viscosity to entropy ratio.}

%\PACS{}
\date{November 2008}

\maketitle

%\tableofcontents
\section{Introduction}
\label{S:Introduction}
Over the recent years, the gauge-string duality has proved to be a valuable tool for understanding strongly coupled conformal field theories. In particular, the shear viscosity $\eta$ of a wide class of strongly-coupled gauge-theory plasmas has been computed. These include the $\mathcal{N}=4$ super-Yang-Mills theory \cite{Policastro:2001yc,Policastro:2002se,Policastro:2002tn}, the theory dual to the near horizon limit of $M$-branes \cite{Herzog:2002fn,Herzog:2003ke}, some non conformal theories \cite{Buchel:2004hw,Parnachev:2005hh,Benincasa:2005iv,Buchel:2005cv,Benincasa:2006ei}, non relativistic theories \cite{Adams:2008wt,Herzog:2008wg,Maldacena:2008wh} and extensions of these theories which include the introduction of a chemical potential \cite{Mas:2006dy,Son:2006em,Maeda:2006by,Saremi:2006ep}, fundamental matter \cite{Mateos:2006yd,Ge:2008ak} and non-commutativity \cite{Landsteiner:2007bd}. In all the cases considered the large 't Hooft, large $N$ value of the ratio of the shear viscosity to entropy density was
\begin{equation}
\label{E:etaovers}
 	\frac{\eta}{\rm s} = \frac{1}{4\pi}.
\end{equation}
General arguments showing that the ratio \eqref{E:etaovers} is universal can be found in \cite{Buchel:2003tz,Kovtun:2004de,Buchel:2004qq}. Extensions of \eqref{E:etaovers} to finite 't Hooft coupling  or $N$ can be found in \cite{Buchel:2004di,Benincasa:2005qc,Benincasa:2006fu,Brigante:2007nu,Dutta:2008gf,Myers:2008yi,Brustein:2008cg,Brustein:2008xx}.  Perhaps of more practical importance is the proximity of \eqref{E:etaovers} to the apparent shear viscosity to entropy ratio of the quark gluon plasma, presumably observed in heavy ion collisions \cite{Romatschke:2007mq,Dusling:2007gi,Luzum:2008cw}. Discussions of the universal behavior of other properties of the quark gluon plasma can be found in \cite{Buchel:2007mf,Gubser:2007ni,Kovtun:2008kx,Iqbal:2008by}.

To leading order in the hydrodynamic approximation, in the absence of generalized forces and when there are no spontaneously broken symmetries, the shear viscosity is the single parameter which is required in order to completely specify the stress-energy tensor in a conformal theory on $\mathbf{R}^{d,1}$.\footnote{
%footnote
As summarized in \cite{Son:2000ht}, a broken symmetry implies an extra velocity field associated with the coherent motion of the condensate. With the appearance of an extra velocity field, one should be able to construct an extra viscous term in the energy momentum tensor. A holographic construction of a two-component fluid can be found in \cite{Herzog:2008he}. }
%endfootnote
Indeed, when the mean free path of the theory is smaller than the typical momentum scale, one can parameterize all the quantities appearing in the hydrodynamic equations by the hydrodynamic variables $u^{\mu}$, $\epsilon$, and $\mu_a$.
The velocity field $u^{\mu}$ is defined as the velocity needed to boost a fluid element to its rest frame, $\epsilon$ is the local energy density in the rest frame of a fluid element, and the $\mu_a$ are local chemical potentials conjugate to charge densities $\rho_a$ also defined relative to the rest frame of a fluid element. With this parameterization, conformal invariance dictates that
\begin{equation}
\label{E:ShearO1}
	\langle T_{\mu\nu} \rangle =
	\frac{1}{d-1} \epsilon(d u_{\mu}u_{\nu} + \eta_{\mu\nu})
	-\eta \sigma_{\mu\nu}+\mathcal{O}(\partial^2)\ ,
\end{equation}
where
\begin{equation}
	\sigma_{\mu\nu} \equiv 2\partial_{<\mu}u_{\nu>}
\end{equation}
and triangular brackets denote a symmetrized, traceless projection:
\begin{equation}
	A_{<\mu\nu>} = \frac{1}{2}P_{\mu}^{\phantom{\mu}\lambda}P_{\nu}^{\phantom{\nu}\sigma}\left(A_{\lambda\sigma}+A_{\sigma\lambda}\right) - \frac{1}{d-1}P_{\mu\nu}P^{\lambda\sigma}A_{\lambda\sigma}
\end{equation}
with
\begin{equation}
 	P_{\mu\nu} = \eta_{\mu\nu}+u_{\mu}u_{\nu}
\end{equation}
a projection onto the space orthogonal to the velocity field.

The $\mathcal{O}(\partial^2)$ in \eqref{E:ShearO1} refers to expressions which are subleading in gradients of the hydrodynamic fields. I.e., they include at least two derivatives of $u^{\mu}$, $\epsilon$ and $\mu_a$.
At second order in the hydrodynamic expansion many more parameters are needed to completely specify the stress energy tensor. We call these parameters second order transport coefficients.
Four of these, $\tau_{\pi}$ and $\lambda_i$ with $i=1,\ldots,3$  were discussed in \cite{Baier:2007ix}. They are associated with the contributions of
\begin{align}
	\Sigma^{(0)}_{\mu\nu} &= {}_{\langle}u^{\lambda}\partial_{\lambda}\sigma_{\mu\nu\rangle}
		+\frac{1}{d-1}\sigma_{\mu\nu}\partial_{\lambda}u^{\lambda}\ ,
	\notag
	\\
	\Sigma_{\mu\nu}^{(1)} &= \sigma_{\langle\mu\lambda}\sigma^{\lambda}_{\phantom{\lambda}\nu\rangle}\ ,
	\qquad
	\Sigma_{\mu\nu}^{(2)} = \sigma_{\langle \mu\lambda}\omega^{\lambda}_{\phantom{\lambda}\nu\rangle}\ ,
	\qquad
	\Sigma_{\mu\nu}^{(3)} = \omega_{\langle \mu\lambda}\omega^{\lambda}_{\phantom{\lambda}\nu\rangle}\ , \end{align}
with
\begin{align}
	\omega_{\mu\nu} = \frac{1}{2}P^{\lambda}_{\phantom{\lambda}\mu}P^{\sigma}_{\phantom{\sigma}\nu}\left(\partial_{\lambda}u_{\sigma}-\partial_{\sigma}u_{\lambda}\right)\ ,
\end{align}
to the second order stress-energy tensor:
\begin{equation}
\label{E:Shear}
	\langle T_{\mu\nu} \rangle =
	\frac{1}{d-1} \epsilon(d u_{\mu}u_{\nu} + \eta_{\mu\nu})
	-\eta \sigma_{\mu\nu}+
	\tau_{\pi}\eta \Sigma^{(0)}_{\mu\nu} + \lambda_i \Sigma^{(i)}_{\mu\nu} + \mathcal{O}(\partial^2).
\end{equation}
The coefficients $\tau_{\pi}$ and $\lambda_i$ were computed for theories whose dual is given by Einstein gravity \cite{Baier:2007ix, Natsuume:2007ty,Bhattacharyya:2008jc, VanRaamsdonk:2008fp,Haack:2008cp,Bhattacharyya:2008mz}, five dimensional Einstein-Maxwell gravity with a single $U(1)$ field \cite{Erdmenger:2008rm,Banerjee:2008th} or five dimensional Einstein gravity in the presence of a dilaton source \cite{Bhattacharyya:2008ji}. In \cite{Erdmenger:2008rm} it was observed that in all these examples, the transport coefficients $\tau_{\pi}$, $\lambda_1$ and $\lambda_2$ satisfy
\begin{equation}
\label{E:Universal2}
	 4 \lambda_1+\lambda_2 = 2 \eta \tau_{\pi}
\end{equation}
when the boundary theory spacetime dimension is larger than three. In three dimensions $\Sigma^{(1)}_{\mu\nu}\equiv 0$, so it makes no sense to define the associated transport coefficient $\lambda_1$. From now on we will always be considering $d>3$.
In this work, we show that \eqref{E:Universal2} holds for a class of strongly coupled, conformal, gauge theory plasmas with conserved global charges which have a bulk dual.
We assume that the bulk dual of the stationary fluid can be described by a stationary and static black hole and that the black hole solution can be extended in the sense of \cite{Bhattacharyya:2008jc}. Further, we show that  $\lambda_2$ is given by
\begin{equation}
\label{E:lambda2val}
%	2 \kappa^2 \lambda_1 &= \frac{1}{2}r_+^{D-3}s(r_+)\\
	\lambda_2 = \kappa^{-2}  \int_{r_+}^{\infty} \frac{\left(r_+^{2 d-2}+(f(x)-1)x^{2 d-2}\right)s(x)}{x^{d+1} f(x)}dx \ ,
\end{equation}
where $s(r)$ and $f(r)$ are given by the metric components of the bulk dual to a static, stationary configuration of the fluid,
\begin{equation}
\label{E:Simple}
	ds^2 =r^2\left(-f(r)dt^2+\sum_{i=1}^{d-1}dx_i^2 \right)
		+2 s(r) drdt,
\end{equation}
$r_+$ is the largest zero of $f(r)$, and $2\kappa^2$ is the $(d+1)$-dimensional gravitational constant of the bulk theory (see below). 
Equation \eqref{E:lambda2val} coincides with what was found in \cite{Bhattacharyya:2008mz} where the case $f(r) = 1-(r_+/r)^d$, $s(r)=1$ was studied.
Note that \eqref{E:Universal2} imposes a constraint amongst the transport coefficients of the theory, whereas \eqref{E:etaovers} relates a transport coefficient (the shear viscosity) to a thermodynamic variable (the entropy density). In that sense \eqref{E:etaovers} and \eqref{E:Universal2} differ. Further discussion of \eqref{E:Universal2} and \eqref{E:lambda2val} and their possible extensions can be found in section \ref{S:Discussion}. The derivation of these two equations can be found in section \ref{S:Main}.

\section{Computing $\langle T_{\mu\nu}\rangle$}
\label{S:Main}

Consider an action of gravity coupled to matter,
\begin{equation}
\label{E:action}
    S =- \frac{1}{2\kappa^2}\int \sqrt{-g} \left(R + \mathcal{L}_M\right) d^{D-1}xdr + S_{\rm CS}.
\end{equation}
The matter Lagrangian may include scalars $\Phi^\alpha$, and Abelian vectors $A^a_M$,
\begin{equation}
\label{E:matter}
    \mathcal{L}_M =-G_{\alpha\beta}(\Phi) \partial_M \Phi^\alpha \partial^M \Phi^\beta
        +V(\Phi) - K_{ab}(\Phi)\left(F_{MN}^{a} F^{MN\,b}\right)
\end{equation}
with $F^a_{MN} = \partial_{M} A^a_{N}-\partial_{N} A^a_{M}$. The indices $a,b$ and $\alpha,\beta$ run over the number of gauge fields and the number of scalar fields, respectively. The indices $M=0,\ldots,D-2,D$ run over the $D$ AdS directions with the $D$ coordinate denoting the radial AdS direction $r$. The indices $\mu,\nu$ which have been introduced in the previous section, run over the $d=D-1$ directions transverse to the radial AdS direction. These will be called the transverse directions for short. The action $S_{\rm CS}$ denotes a possible Chern-Simons term.
The Einstein equations which follow from the action \eqref{E:action} are
\begin{equation}
\label{E:Einstein}
    R_{MN} = \mathfrak{T}_{MN}+g_{MN} \mathfrak{T}\ ,
\end{equation}
where
\begin{subequations}
\label{E:bulkTmn}
\begin{align}
\label{E:TauMN}
	\mathfrak{T}_{MN} &= G_{\alpha\beta} \partial_M \Phi^{\alpha} \partial_N \Phi^{\beta}
		+ 2 K_{ab}F_{MA}^a F^{b \phantom{N}A}_{\phantom{b}N}\ , \\
	\mathfrak{T} & = -\frac{1}{(D-2)}\left(V(\Phi) +  K_{ab}(\Phi)F_{MN}^{a} F^{MN\,b}\right).
\end{align}
\end{subequations}
In \eqref{E:Einstein} the matter stress-energy tensor has been decomposed into a term $g_{MN}\mathfrak{T}$, whose index structure is proportional to $g_{MN}$, and another term $\mathfrak{T}_{MN}$ whose index structure comes from the gauge fields, their derivatives, and derivatives of the scalar fields.

Since we are interested in a thermal state in the dual theory, we consider black hole solutions to \eqref{E:Einstein}. Let us assume that the equations of motion allow for a translation and rotation invariant (in the spatial transverse directions) time-independent, asymptotically AdS black brane solution. In the Eddington-Finkelstein coordinate system, the most general such solution takes the form \eqref{E:Simple}. 
The asymptotically AdS boundary is located at $r \to \infty$ where $f (\infty)=1$ and $s(\infty)=1$.
We denote the largest zero of $f(r)$ by $r_+$. This is the location of the outermost horizon of the black hole. 
Assuming that the gauge and scalar fields are also time-independent and invariant under translations and rotations (i.e., $\Phi^{\alpha} = \Phi^{\alpha}(r)$, $A_i^a = 0$, $A_0^a = 
A_0^a(r)$), we find that
\begin{equation}
\label{E:Tijis0}
    \mathfrak{T}_{ij}=0\ , \quad \mathfrak{T}_{0i}=0\ ,
\end{equation}
where the indices $i$ and $j$ run over the transverse, spatial directions. To make contact with the notation in the rest of the paper we denote the components of $\mathfrak{T}_{MN}$ and $\mathfrak{T}$ by
$\mathfrak{T}_{MN}^{(0)}$, $\mathfrak{T}^{(0)}$.

Inserting \eqref{E:Tijis0} into \eqref{E:Einstein} we can determine $f(r)$ and $s(r)$ in terms of $\mathfrak{T}^{(0)}$ and $\mathfrak{T}^{(0)}_{DD}$ through the $M=i,N=i$ and $M=D,N=D$ Einstein equations. We find that
\begin{subequations}
\label{E:EOM0}
\begin{align}
\label{E:fsol}
	f(r) &= -\frac{s(r)}{r^{D-1}}\int^r x^{D-2}\mathfrak{T}^{(0)}(x) s(x) dx +\frac{\left(\epsilon-\epsilon_0\right)s(r)}{2\kappa^2(D-2)r^{D-1}}\, ,\\
\label{E:Ssol}
	s(r) &= e^{-\frac{1}{D-2}\int_r^{\infty} x \mathfrak{T}^{(0)}_{DD}(x) dx}.
\end{align}
\end{subequations}
One can show that these solutions are consistent with the remaining Einstein equations as long as the bulk energy momentum tensor is conserved, $D_{M}T^{MN}=0$.
The extra integration constant, ${(\epsilon-\epsilon_0)}/(2\kappa^2(D-2))$, has been chosen to be the coefficient of the $r^{-(D-1)}$ term in a series expansion of $f(r)$ near the asymptotically AdS boundary. The reason for this notation will become clear shortly.
The requirement that the boundary is asymptotically AdS has fixed the upper end of the integral in \eqref{E:Ssol}, ensuring that $s(\infty)=1$. It also implies that $\mathfrak{T}^{(0)}(\infty) = -(D-1)$ in order that $f(\infty)=1$ be satisfied. 

In this work we will further require that
\begin{subequations}
\label{E:fsAasymptotics}
\begin{equation}
\label{E:fands}
	f(r)-1 = \mathcal{O}(r^{-(D-1)})\ ,
	\quad
	s(r)-1 = \mathcal{O}(r^{-(D-1)}).
\end{equation}
The reason is as follows. The bulk equations of motion specify the metric up to two sets of integration constants. The first set specifies the value of the metric at the asymptotically AdS boundary which in turn specifies the metric of the dual CFT. The second set of integration constants may be associated with the $\mathcal{O}(r^{-(D-1)})$ coefficients  of a near boundary series expansion of the bulk-metric (as specified above) and are associated with the energy momentum tensor of the dual CFT. Once these two sets of integration constants are specified, the full bulk metric is completely fixed. Its near boundary Taylor series can be determined by iteratively solving the equations of motion. Generically, for odd $D$ a non-flat boundary metric induces  logarithmic terms in the near boundary series expansion of the bulk metric. At order $\mathcal{O}(r^{-(D-1)}\ln r)$ these logarithms are associated with conformal anomalies which render the trace of the boundary theory stress-energy tensor non-vanishing (as expected when the boundary theory geometry is non-flat). But it is not only the boundary metric which can induce such logarithmic terms. In odd dimensions, any time the bulk stress energy tensor generates an $\mathcal{O}(r^{-2m})$ term in the metric, with $m< (D-1)/2$, there will also be an $\mathcal{O}(r^{-(D-1)}\ln r)$ term which implies that the state in the boundary theory is not conformally invariant. An explicit example of such behavior can be found in \cite{Gubser:2008vz}. We will also restrict ourselves to solutions for which
\begin{equation}
	A^a_{\mu} = \mathcal{O}(r^{-(D-3)})\delta_{\mu}^0
\end{equation}
\end{subequations}
and will set the sources dual to the scalars to zero.
An example of an asymptotically AdS background satisfying \eqref{E:fsAasymptotics} is given in \cite{Behrndt:1998jd,Cvetic:1999ne}.\footnote{Note that we do not require the gauge field to vanish at the future horizon and therefore it is likely that it diverges at the past horizon. In fact, it is likely that the whole perturbative solution diverges at the past horizon because generic solutions of viscous fluid dynamics are not expected to be regular in the infinite past. We thank A. Karch, D. Son, and especially R. Loganayagam for clarifying this point.}
We discuss the effects of loosening the restriction \eqref{E:fsAasymptotics} in section \ref{S:Discussion}.
 
We may now use the prescription of \cite{Witten:1998qj,Gubser:1998bc,Balasubramanian:1999re,deHaro:2000xn}, adapted to Eddington-Finkelstein coordinates \cite{Haack:2008cp}, to convert the metric \eqref{E:Simple} to the boundary theory stress-tensor $\langle T_{\mu\nu} \rangle$. We find that
\begin{equation}
\label{E:Tmnstatic}
	\langle T_{\mu\nu} \rangle  = 
	\frac{\epsilon-\epsilon_0}{d-1} \left(d\delta_{\mu}^0\delta_{\nu}^0+\eta_{\mu\nu}\right)
	+\frac{\epsilon_0}{d-1} \left(d\delta_{\mu}^0\delta_{\nu}^0+\eta_{\mu\nu}\right).
\end{equation}
The first term in \eqref{E:Tmnstatic} represents the component of the energy momentum tensor associated with the extrinsic curvature of the near boundary metric of the black hole and a boundary cosmological constant counterterm to the bulk action. The second term in \eqref{E:Tmnstatic} represents extra contributions arising from the matter content of the theory, and from possible matter counterterms when holographically renormalizing the bulk action \cite{Skenderis:2002wp}. Since we have assumed that the theory is exactly conformal, these extra contributions can be determined up to an overall factor, $\epsilon_0$.

So far, we have considered a stationary and static black hole. Following the philosophy of \cite{Bhattacharyya:2008jc}, we assume that the black hole solution \eqref{E:Simple} can be extended to capture the fluid dynamical nature of its boundary dual. We start by boosting the transverse coordinates by a constant velocity $u^{\mu}$. After this boost, the metric \eqref{E:Simple} takes the form
\begin{equation}
\label{E:Boosted}
	ds^2 = {r^2}
	\left(-f(r)u_{\mu}u_{\nu}dx^{\mu}dx^{\nu} + P_{\mu\nu}dx^{\mu}dx^{\nu}\right) - 2 s(r) u_{\mu} dx^{\mu} dr.
\end{equation}
Clearly, \eqref{E:Boosted} is still a solution to the Einstein equations since
it has been obtained from the solution \eqref{E:Simple} by a coordinate transformation.
The stress energy tensor of the boundary theory dual to \eqref{E:Boosted} is the boosted version of the stress energy tensor \eqref{E:Tmnstatic}
\begin{equation}
	\langle T_{\mu\nu} \rangle  = 
	\frac{\epsilon}{d-1} \left(du_{\mu}u_{\nu}+\eta_{\mu\nu}\right).
\end{equation}
As explained earlier, in the boundary theory, the hydrodynamic approximation involves allowing the velocity field, energy density and chemical potentials to vary slowly in the transverse directions. Thus, to obtain the gravity dual of hydrodynamics, we allow the bulk quantities, $u^{\mu}$, $\epsilon$ and the various charges to slowly vary with $x^{\mu}$. Once we do that, \eqref{E:Boosted} will no longer be a solution to the equations of motion. Therefore, we add a correction $g^{(1)}_{\mu\nu}$ to the metric \eqref{E:Boosted}, linear in derivatives of the velocity fields, energy density, and whatever chemical potentials are present. Similar corrections, $\Phi^{\alpha\,(1)}$ and $A^{a\,(1)}_{\mu}$, are added to the scalars and gauge fields. The explicit values of $g^{(1)}_{\mu\nu}$, $\Phi^{\alpha\,(1)}$ and $A^{a\,(1)}_{\mu}$ are determined by solving the Einstein equations \eqref{E:Einstein} together with the equations of motion for the matter fields.
Once the corrections which are linear in the derivatives are obtained, one can continue this procedure to obtain two-derivative corrections  $ g^{(2)}_{\mu\nu}$, $\Phi^{\alpha\,(2)}$ and $A^{a\,(2)}_{\mu}$, then three-derivative corrections, etc. Such a perturbation scheme has been shown to work (up to the two derivative level) in the case of Einstein gravity in various dimensions \cite{Baier:2007ix, Natsuume:2007ty,Bhattacharyya:2008jc, VanRaamsdonk:2008fp,Haack:2008cp,Bhattacharyya:2008mz}, in the case of Einstein theory coupled to a dilaton \cite{Bhattacharyya:2008ji} and Einstein-Maxwell gravity in five dimensions with and without a Chern-Simons term \cite{Erdmenger:2008rm,Banerjee:2008th,Hur:2008tq} . 

After carrying out this perturbative scheme, one will end up with a (perturbative) solution to the Einstein equations dual to a non trivial fluid flow in the boundary theory. In \cite{Bhattacharyya:2008ji} it has been shown that conformal invariance of the boundary theory implies that the bulk metric must take the form
\begin{multline}
\label{E:Cinvariantmetric}
	ds^2 = -2 S  \left(dr +r \left(u^{\lambda}\partial_{\lambda}u_{\nu} - \frac{1}{d-1}\partial_{\lambda}u^{\lambda}u_{\nu}\right)dx^{\nu}\right)u_{\mu} dx^{\mu}\\
		+r^2 K u_{\mu}u_{\nu}dx^{\mu}dx^{\nu}
		+r^2 P_{\mu\nu}dx^{\mu}dx^{\nu}
		+r^2 J_{\lambda} \left(P^{\lambda}_{\phantom{\lambda}\mu}u_{\nu}+P^{\lambda}_{\phantom{\lambda}\nu}u_{\mu}\right)dx^{\mu}dx^{\nu}
		+r^2 \pi_{\mu\nu} dx^{\mu}dx^{\nu},
\end{multline}
where $S(r)$, $K(r)$, $J_{\mu}(r)$ and $\pi_{\mu\nu}(r)$ are invariant under the boundary theory Weyl-rescaling $\eta_{\mu\nu} \to e^{-2\chi}\eta_{\mu\nu}$ and depend implicitly on $x^{\mu}$.\footnote{
In \cite{Bhattacharyya:2008ji} the bulk form of the metric was slightly more general than \eqref{E:Cinvariantmetric} in that the coefficient of the $r^2 P_{\mu\nu}$ term was an additional Weyl invariant scalar. Here we have chosen a gauge where this extra scalar has been set to one.}
In addition, the traceless tensor $\pi_{\mu\nu}$ is orthogonal to the velocity field $u^{\mu}$.  
If we set
\begin{equation}
	S(r) = s(r)\ , \quad
	K(r) = -f(r)\ , \quad
	J_{\mu}(r) = 0\ , \quad
	\pi_{\mu\nu}(r)=0 \ ,
\end{equation}
and set $u^{\mu}$ to a constant,
we recover our boosted stationary solution \eqref{E:Boosted}.
Information on the viscous corrections to the boundary theory energy momentum tensor is contained in $\pi_{\mu\nu}$. This follows from the mapping of the bulk metric to the boundary stress energy tensor. In our notation (see \cite{Haack:2008cp}),
\begin{equation}
\label{E:BtB}
	\langle T_{\mu\nu} \rangle =
		\frac{\epsilon}{d-1} \left(du_{\mu}u_{\nu}+\eta_{\mu\nu}\right)
		-\frac{\kappa^{-2}}{2}{\rm Finite}\left[\lim_{r\to\infty}r^{D}\partial_r \pi_{\mu\nu}(r)\right]
		+\ldots \ ,
\end{equation}
where ${\rm Finite}[\lim_{r\to\infty}B(r)]$ means the finite part of $B(r)$ as $r$ is taken to infinity. 
The ellipses in \eqref{E:BtB} represent corrections to the Brown York stress tensor coming from holographic renormalization and from the matter content of the theory. 
Thus, to find the various transport coefficients associated with the boundary theory fluid, we need to compute $\pi_{\mu\nu}$ and deal with the various corrections to $\langle T_{\mu\nu} \rangle$ coming from holographic renormalization and from varying the matter action with respect to the boundary metric.

As a warmup, let us compute the shear viscosity of gauge theory plasmas whose dual bulk action is of the form \eqref{E:action}. For this we need to find the stress energy tensor of the plasma to first order in the hydrodynamic approximation. Equivalently, in view of \eqref{E:BtB}, we need to find $\pi_{\mu\nu}$ to first order. As stated earlier, we denote this term $\pi^{(1)}_{\mu\nu}$ and in what follows we will use the notation $B^{(n)}$ to specify the $n$'th order component in a derivative expansion of the field $B$.

As pointed out in \cite{Bhattacharyya:2008jc}, a derivative expansion of the Einstein equations results in a differential equation in the radial coordinate only. This implies that it is sufficient to solve the Einstein equations in the neighborhood of an arbitrary point $x_0^{\mu}$ in the transverse directions for all values of $r$ and then extend the solution to the entire manifold using Lorentz invariance. A further simplification can be made if we choose a coordinate system where 
\begin{equation}
	u^{\mu}(x_0)= (1,  0,  \ldots, 0 ). 
\end{equation}	
This is a simplification since in the neighborhood of $x_0^{\mu}$, the Einstein equations naturally decompose into tensor, vector and scalar modes of the $SO(D-2)$ symmetry which leaves $u^{\mu}(x_0)$ invariant. The equations of motion for the tensor modes, $\pi_{ij}(x_0)$, can be read off from the $M=i$, $N=j\neq i$ components of the Einstein equations \eqref{E:Einstein} expanded around $x_0^{\mu}$,
\begin{equation}
\label{E:Tensor1}
    R_{ij}^{(1)} = g^{(1)}_{ij} \mathfrak{T}^{(0)}+g_{ij}^{(0)}\mathfrak{T}^{(1)}+\mathfrak{T}_{ij}^{(1)}.
\end{equation}
Working out $R_{ij}^{(1)}$ explicitly, using $g_{ij}^{(0)} = \mathfrak{T}_{ij}^{(1)} = 0$ for $i \neq j$ and using \eqref{E:fsol} we find that \eqref{E:Tensor1} reads
\begin{equation}
\label{E:pi1EOM}
	\left(\frac{r^{D}f}{s}  \pi_{ij}^{(1)\,\prime} \right)^{\prime} = - (D-2)r^{D-3} \sigma_{ij}\ ,
\end{equation}
where a prime denotes a derivative in the radial direction.
After extending the solution to \eqref{E:pi1EOM} to the entire manifold it takes the form
\begin{equation}
\label{E:pival}
	\pi_{\mu\nu}^{(1)}(r)  =  \sigma_{\mu\nu} F(r)
\end{equation}
with
\begin{equation}
\label{E:Fval}
	F(r) = \int_r^{\infty} \frac{(x^{D-2}-r_+^{D-2})s(x)}{x^D f(x)}dx.
\end{equation}
While $\pi_{\mu\nu}(r)$ can only be written down implicitly, we can use \eqref{E:Fval} to obtain
\begin{equation}
\label{E:etaval}
	{\rm Finite}\left[
		\lim_{r \to \infty} r^{D} \partial_r \pi^{(1)}_{\mu\nu}(r)
		\right]  =  r_+^{D-2}.
\end{equation}

In order to use  \eqref{E:BtB}  to evaluate the boundary theory energy momentum tensor, we need access to the extra terms in \eqref{E:BtB} arising from holographic renormalization and the matter fields.
By using \eqref{E:action} and \eqref{E:fsAasymptotics} and a simple power counting argument, one can show that the gauge fields do not contribute to such terms. We elaborate on this argument in appendix \ref{A:HR}. We conclude that only terms involving derivatives of the scalars can contribute to the boundary theory energy momentum tensor.
As discussed in the introduction, for the theories we have in mind, such expressions can not contribute to $\langle T_{\mu\nu} \rangle$ at the one derivative level. Thus,
\begin{equation}
	\eta =  \frac{r_+^{D-2}}{2\kappa^2}.
\end{equation}
To compare this with \eqref{E:etaovers}, one can compute the entropy density of the boundary theory fluid from the Bekenstein-Hawking formula:
\begin{equation}
	{\rm s} = \frac{4\pi r_+^{D-2}}{2\kappa^2}.
\end{equation}
The ratio $\eta/{\rm s}$ agrees with \eqref{E:etaovers}. For the class of theories we are considering, this result was previously obtained in \cite{Benincasa:2006fu}.

We proceed to compute $\pi_{\mu\nu}^{(2)}$, the second order contribution to $\pi_{\mu\nu}$. The computation is similar to the one that lead to \eqref{E:pival}. At second order in the derivative expansion, $\pi_{\mu\nu}^{(2)}$ can be computed from the $i \neq j$ components of the Einstein equation expanded around $x_0^{\mu}$
\begin{equation}
	R^{(2)}_{ij} = \mathfrak{T}^{(0)} g^{(2)}_{ij} + \mathfrak{T}^{(1)} g^{(1)}_{ij} + \mathfrak{T}^{(2)}_{ij}.
\end{equation}
Writing this explicitly in terms of the metric components, we find
\begin{multline}
\label{E:EOM2nd}
	\left(\frac{r^{D}f}{s}\pi_{ij}^{(2)\,\prime}\right)^{\prime}= 
		\sum_{n=0,\ldots,3}P_n \Sigma^{(n)}_{ij} +
		\sum_{n=1,2}\overline{P}_n \overline{\Sigma}^{(n)}_{ij}
		+\left(\mathcal{T} -2 r^{D-2} s F \, \mathfrak{T}^{(1)} \right)\sigma_{ij} -2 r^{D-4} s \,\mathfrak{T}^{(2)}_{ij}\\
		-2 \left(r^{D-2} {\phantom{\big |}\partial}_{\langle i}J^{(1)}_{j\rangle}\right)^{\prime}
		-2(D-2) r^{D-3} J^{(1)}_{\langle i}\partial_0 u {\phantom{\big|}\negthickspace}_{j \rangle}
		- r^D \frac{J^{(1)\,\prime}_{\langle i}J^{(1)\,\prime}_{j\rangle}}{s}
		+\frac{2r^{D+1}(s r^{-2})^{\prime}J^{(1)\,\prime}_{\langle i}\partial_0 u {\phantom{\big|}\negthickspace}_{j \rangle}}{s}\\
		-2 r^{D-4}\partial_{\langle i}\partial_{j \rangle}s
		+\frac{r^{D-4}\partial_{\langle i}s\partial_{j \rangle}s}{s}
		+2 r \left(r^{D-4} \partial_{\langle i}s \partial_0 u_{j \rangle}\right)^{\prime}
		.
\end{multline}
The quantities $P_n$, $\overline{P}_n$ and $\mathcal{T}$ are functions of $r$, $s(r)$, $f(r)$, $S^{(1)}(r)$, $K^{(1)}(r)$, $F(r)$ and their derivatives. We refrain from writing them out explicitly. The $\overline{\Sigma}^{(n)}_{\mu\nu}$'s are given by
\begin{equation}
	\overline{\Sigma}^{(1)}_{\mu\nu} = \partial_{\sigma} u_{\langle \mu} \partial^{\sigma} u_{\nu \rangle}\ , \quad
	\overline{\Sigma}^{(2)}_{\mu\nu} = u^{\sigma}\partial_{\langle \mu}\partial_{\sigma } u_{\nu \rangle}.
\end{equation}
The superscripts of the $\Sigma$'s and $\overline{\Sigma}$'s are a means to enumerate these components and are not related to the hydrodynamic expansion. 
$J_{i}^{(1)}$ is the first order correction to $J_{\mu}$ defined in \eqref{E:Cinvariantmetric}, expanded around $x_0^{\mu}$. It can be determined in terms of the vector components of $\mathfrak{T}_{MN}^{(1)}$ by solving the first order vector mode equation equivalent of \eqref{E:Tensor1}. 
The expressions for $S^{(1)}$ and $K^{(1)}$ can also be determined by solving the first order scalar equations. It is straightforward to show that $S^{(1)}=0$. In \eqref{E:EOM2nd} terms proportional to $x_{\mu}$ have been omitted.

Notice that the index structure of $\pi_{ij}^{(2)}$ is inherited from the index structure of the terms on the right hand side of \eqref{E:EOM2nd}. Further, the index structure of the boundary theory stress energy tensor is equal to that of $\pi_{ij}$ plus any terms coming from the matter fields or counterterms from holographic renormalization. Before proceeding, let us take a moment to characterize the possible terms which can contribute to $\pi_{ij}^{(2)}$.  Recall that $\pi_{ij}$ is a Weyl invariant traceless symmetric tensor orthogonal to the velocity field. At second order in the derivative expansion there are a finite number of combinations of derivatives of $\epsilon$, $u_{\mu}$ and $\mu_a$  which can contribute to $\pi_{ij}^{(2)}$. If we consider terms composed only of derivatives of $u_{\mu}$ and $\epsilon$ and use energy-momentum conservation, $\partial^{\mu}\langle T_{\mu\nu} \rangle = 0$, or 
\begin{subequations}
\label{E:Conservation}
\begin{align}
\label{E:epsilonspace}
	P_{\mu}^{\phantom{\mu}\nu}\partial_{\nu} \epsilon = - \epsilon d u^{\nu} \partial_{\nu} u_{\mu} + \mathcal{O}(\partial^2)\, ,\\
\label{E:epsilontime}
	u^{\mu} \partial_{\mu} \epsilon = -\epsilon \frac{d}{d-1} \partial_{\nu}u^{\nu} + \mathcal{O}(\partial^2),
\end{align}
then the only Weyl invariant, even-parity terms which we can construct are $\Sigma^{(n)}_{\mu\nu}$ with $n=0,\ldots , 3$. If we loosen the restriction on Weyl invariance there are two more possible terms: $\overline{\Sigma}^{(1)}_{\mu\nu}$ and  $\overline{\Sigma}^{(2)}_{\mu\nu}$ \cite{Bhattacharyya:2008jc}. Allowing for pseudo-tensors, or terms which involve $\mu_a$ and their derivatives, will lead to many more possibilities. We denote these other terms by $\Sigma^{(n)}_{\mu\nu}$ with $n \geq 4$. We exclude from this list expressions which are equivalent under current conservation
$\partial^{\mu}\langle J^a_{\mu} \rangle = 0$, or
\begin{equation}
	u^{\sigma}\partial_{\sigma}\rho_a = - \rho_a \partial_{\sigma}u^{\sigma} + \mathcal{O}(\partial^2).
\end{equation}
\end{subequations}
In what follows, we will rewrite all the terms on the right hand side of \eqref{E:EOM2nd} in terms of the basis $\Sigma^{(n)}_{\mu\nu}$ expanded around $x_0^{\mu}$. Our assumption regarding conformal invariance dictates that the coefficients of the non Weyl-invariant expressions will eventually vanish. However, as we will see below, it will be useful to keep track of the coefficients of the $\overline{\Sigma}^{(n)}_{\mu\nu}$ terms.

Let us start with the $\left(\mathcal{T} -2 r^{D-2} s F \mathfrak{T}^{(1)} \right)\sigma_{ij}$ term in the first line of \eqref{E:EOM2nd}. Using \eqref{E:Conservation}, at the one derivative level we find that the only scalar which can be constructed out of $u_{\mu}$, $\mu_a$ and $\epsilon$  is $\partial_{\mu} u^{\mu}$. 
Thus,
\begin{equation}
\label{E:scalarthetas}
	\left(\mathcal{T}(r) -2 r^{D-2} s F \mathfrak{T}^{(1)}(r) \right) = \theta_1(r) \partial_{\mu} u^{\mu}.
\end{equation}

Next consider the various derivatives of $J^{(1)}_{i}$. The Weyl invariant vector or pseudo-vector $J^{(1)}_{\nu}$ is constructed from $u^{\mu}$, $\epsilon$ and $\mu_a$ and one derivative. From \eqref{E:Cinvariantmetric} we see that only the component of $J^{(1)}_{\nu}$ which is orthogonal to the velocity field will enter the metric. Thus, $J^{(1)}_{\nu}$ must be composed of
\begin{subequations}
\label{E:O1vecs}
\begin{equation}
	\partial_{\nu} \frac{\mu_a^d}{\epsilon},
	\quad
	\partial_{\nu} \frac{\mu_a}{\mu_b}
\end{equation}
and, in $d=4$, 
\begin{equation}
	\ell_{\mu} = \epsilon_{\mu\nu\lambda\sigma}u^{\nu}\partial^{\lambda}u^{\sigma}.
\end{equation}
\end{subequations}
Using \eqref{E:epsilonspace}, we find that the expressions involving $J_i^{(1)}$ in \eqref{E:EOM2nd} must take the form
\begin{equation}
\label{E:vectorthetas}
	\theta_2(r) \partial_0 u_{\langle j} \partial_0 u_{i \rangle}
	+\theta_3(r) \partial_{\langle i} \left(u^{\sigma} \partial_{\sigma} u_{j \rangle}\right)
	+\ldots \ ,
\end{equation}
where the ellipses denote expressions which can not possibly contribute to $\Sigma_{ij}^{(n)}$ with $n \leq 2$. (Note that $\ell_{\mu}$ is a pseudo-vector and that $\ell_{\langle \mu} \ell_{\nu \rangle} = 4 \Sigma^{(3)}_{\mu\nu}$.)
Like $J_{\mu}^{(1)}$, $s$ is also Weyl invariant. Therefore, the third line in \eqref{E:EOM2nd} will also take the form \eqref{E:vectorthetas}.

We are left with the $\mathfrak{T}^{(2)}_{ij}$ term in the first line of \eqref{E:EOM2nd}, defined in \eqref{E:TauMN}. This term has two components, one involving the gauge fields, and the other involving the scalars. Using the same argument as in the last paragraph, we find that the scalars can contribute to the $\theta_2$ term in \eqref{E:vectorthetas}.  
To see what the contribution of the $F^a_{MA}F^{b\phantom{N}A}_{\phantom{b}N}$ term in \eqref{E:TauMN} looks like, we expand the gauge field $A^a_{\mu}$ in a derivative expansion, i.e.\
\begin{equation}
\label{E:Amuseries}
	A^a_{\mu}  = c^{a(0)} u_{\mu} + \left(P^{\nu}_{\mu}a^{a\,(1)}_{\nu}+ c^{a\,(1)}u_{\mu}\right) + \ldots\  .
\end{equation}
The first term on the right hand side of \eqref{E:Amuseries}, $c^{a(0)}$, is the zero order stationary solution boosted to a velocity $u_{\mu}$. The second expression, $\left(P^{\nu}_{\mu}a^{a\,(1)}_{\nu}+ c^{a\,(1)}u_{\mu}\right)$, is the first order correction to $A^a_{\mu}$ separated into a contribution orthogonal to the velocity field and a contribution parallel to it. As was the case for $\langle T_{\mu\nu} \rangle$, in order for the boundary current $\langle J^{a}_{\mu} \rangle$ to transform homogeneously under Weyl rescalings, we need that $a^{a\,(1)}_{\nu}$ be a Weyl invariant vector. Plugging \eqref{E:Amuseries} into $F^a_{MA}F^{b\phantom{N}A}_{\phantom{b}N}$ and expanding around $x_0^{\mu}$, we find that for $i \neq j$
\begin{multline}
	\left(F^a_{i A}F^{b\phantom{N}A}_{\phantom{b} j}\right)^{(2)} 
	= -4 r^{-2} c^{a(0)} c^{b(0)} \Sigma^{(3)}_{ij}
	+\frac{r^{2}f}{s^2} a^{a\,(1)\,\prime}_i a^{b\,(1)\,\prime}_j
	\\
	+\frac{\left(\partial_i c^{a(0)} + c^{a(0)} \partial_0 u_i\right) a^{b\,(1)\,\prime}_j}{s} 
	+\frac{\left(\partial_j c^{b(0)} + c^{b(0)} \partial_0 u_j\right) a^{a\,(1)\,\prime}_i}{s} .
\end{multline}
We have omitted terms which depend explicitly on $x^{\mu}$.
The first expression contributes only to $\Sigma^{(3)}_{ij}$ and the remaining terms can only contribute to $\theta_2$ in \eqref{E:vectorthetas}, or to $\Sigma_{\mu\nu}^{(n)}$ with $n \geq 3$.

We can summarize the previous discussion by recasting \eqref{E:EOM2nd} into
\begin{multline}
\label{E:EOM2ndshort}
	\left(\frac{r^{D}f\,\pi_{ij}^{(2)\,\prime}}{s}\right)^{\prime}= 
		\sum_{n=0,1,2}P_n\Sigma^{(n)}_{ij} +
		\sum_{n=1,2}\overline{P}_n \overline{\Sigma}^{(n)}_{ij} \\
		+\theta_1 \partial_{\mu}u^{\mu} \sigma_{ij}
		+\theta_2 \partial_0 u_{\langle j} \partial_0 u_{i \rangle}
		+\theta_3 \partial_{\langle i} \left( u^{\sigma} \partial_{\sigma} u_{j \rangle}\right)
		+\ldots\ ,
\end{multline}
where the ellipses represent expressions which do not contribute to $\overline{\Sigma}^{(n)}_{\mu\nu}$ and $\Sigma^{(n)}_{\mu\nu}$ with $n \leq 2$.
The last three terms in \eqref{E:EOM2ndshort} are symmetric traceless tensors composed of the velocity fields and their derivatives. Therefore, they may be expanded in terms of the basis $\{ \Sigma_{ij}^{(0)}, \ldots , \Sigma_{ij}^{(3)} , \overline{\Sigma}_{ij}^{(1)} , \overline{\Sigma}_{ij}^{(2)} \}$. 
Doing so, equation \eqref{E:EOM2ndshort} takes the form
\begin{equation}
\label{E:EOM2ndshortest}
		\left(\frac{r^{D}f }{s}  \pi_{ij}^{(2)\,\prime}\right)^{\prime}  
		= \sum_{n=0,1,2} \Lambda_n \Sigma^{(n)}_{ij}
		 +\sum_{n=1,2} \overline{\Lambda}_n \overline{\Sigma}^{(n)}_{ij} 
		 +\ldots
\end{equation}
with
\begin{subequations}
\label{E:Lambdaterms}
\begin{alignat}{4}
		\Lambda_0(r)&=P_0(r)
			&{}+\frac{3(D-2)}{D+4}&\theta_1(r)
			&{}+\frac{3}{D+4} &\theta_2(r)
			&{}+\frac{3}{D+4}&\theta_3(r)\ , \\
		\Lambda_1(r)&= P_1(r)
		&{}-\frac{3(D-2)}{2(D+4)}&\theta_1(r)
		&{}+\frac{D-2}{4(D+4)} &\theta_2(r) 
		&{}+ \frac{D-2}{4(D+4)}&\theta_3(r)\ , \\
		\Lambda_2(r)&=
			P_2(r)
			&{}+\frac{6(D-2)}{D+4} &\theta_1(r)
			&{}- \frac{D-2}{D+4}&\theta_2(r) 
			&{}+ \frac{6}{D+4}& \theta_3(r)\ , \\
		\overline{\Lambda}_1(r) &=
		\overline{P}_1(r)
		 &{}+\frac{6(D-2)}{D+4} &\theta_1(r)
		&{}-\frac{D-2}{D+4} &\theta_2(r)
		&{}+\frac{6}{D+4} &\theta_3(r)\ , \\
	\overline{\Lambda}_2(r) &=
	\overline{P}_2(r) 
	&{}-\frac{6(D-2)}{D+4} &\theta_1(r)
	&{}-\frac{6}{D+4} &\theta_2(r)
	&{}+ \frac{D-2}{D+4} &\theta_3(r) \ .
\end{alignat}
\end{subequations}

So far \eqref{E:EOM2ndshortest} shows us the contribution of $\pi_{ij}^{(2)}$ to the boundary theory stress tensor. Further contributions to $\langle T_{\mu\nu} \rangle$ will come from the matter action and from holographic counterterms. We have argued previously that these can only come from the scalar sector of the theory. The only possible contributions from this sector which are relevant to our analysis are transverse, traceless derivatives of the Weyl invariant quantity $\mu^d_a/\epsilon$.  We can use the same logic that brought us from \eqref{E:O1vecs} to \eqref{E:vectorthetas} to argue that these contributions will be proportional to $\partial_0 u_{\langle j} \partial_0 u_{i \rangle}$ and $\partial_{\langle i} \left(u^{\sigma} \partial_{\sigma} u_{j \rangle}\right)$. Now, since the $\theta_i$ terms in \eqref{E:Lambdaterms} are (and will remain) unspecified, we can work with an effective $\theta_2$ and $\theta_3$ which capture the contributions of these extra terms from the scalar sector. In what follows we assume that such a procedure has been carried out, and then the sole contribution to $\tau_{\pi}$, $\lambda_1$ and $\lambda_2$ comes from the $\Lambda_i$, $i \leq 2$ defined in \eqref{E:Lambdaterms} with unspecified $\theta_i$.

We are down to solving \eqref{E:EOM2ndshortest}. Currently, this is made difficult due to the unknown functions $\theta_i$.  
Fortunately, we can do without them. Recall  that the coefficients of the $\overline{\Sigma}^{(n)}_{ij}$ terms must vanish if the boundary theory energy momentum tensor is to be conformally invariant. This implies that
\begin{align}
\begin{split}
\label{E:thetas}
	\theta_1 &= -\frac{1}{D-2} \overline{P}_1 +  \frac{1}{6} \overline{P}_2 + \frac{D-8}{6(D-2)}\theta_3\ , \\
	\theta_2 &= \overline{P}_1 + \overline{P}_2 + \theta_3.
\end{split}
\end{align}
Substituting \eqref{E:thetas} into \eqref{E:EOM2ndshortest} and inserting the explicit values for the $P_n$'s and $\overline{P}_n$'s, we find that
\begin{subequations}
\label{E:Lambdas}
\begin{align}
	\Lambda_0(r) &=r \left(r^{D-4}s(r)\right)^{\prime}-2 r^{\frac{D-2}{2}}\left(r^{\frac{D-2}{2}}F(r)\right)^{\prime} +\frac{1}{2}\theta_3(r)\ , \\	
	\Lambda_1(r) &= \left(\frac{1}{2}  r^{D-3}s(r)-\left(r^{D-2}-r_+^{D-2}\right)F(r)\right)^{\prime}
	+\frac{1}{4}\theta_3(r)\ , \\
\label{E:Lambda2}
	\Lambda_2(r) &= -2 r^{D-4}s(r)-4 r^{\frac{D-2}{2}}\left(r^{\frac{D-2}{2}}F(r)\right)^{\prime}\ ,
\end{align}
\end{subequations}
and, of course, the $\overline{\Lambda}_n$'s vanish by construction.
From \eqref{E:Lambdas} it follows that
\begin{equation}
\label{E:Lambda}	
	4 \Lambda_1+ \Lambda_2 - 2 \Lambda_0 = 4 \left(\left(r^{D-2}-r_+^{D-2}\right)F(r)\right)^{\prime}
\end{equation}
and using \eqref{E:BtB} and \eqref{E:Shear},
we obtain the proposed relation \eqref{E:Universal2}. 
Since $\Lambda_2$ in \eqref{E:Lambda2} does not depend on $\theta_3$ we may use \eqref{E:BtB} and \eqref{E:Shear} to evaluate $\lambda_2$ explicitly. The result is given in \eqref{E:lambda2val}. 

\section{Discussion}
\label{S:Discussion}
In section \eqref{S:Main} we derived our main result, which we reproduce here for convenience,
\begin{equation}
\tag{\ref{E:Universal2}}
	4 \lambda_1 + \lambda_2 = 2 \eta \tau_{\pi} .
\end{equation}
The transport coefficients $\lambda_2$, $\lambda_1$ and $\tau_{\pi}$ are associated with $\Sigma^{(2)}_{\mu\nu}$, $\Sigma^{(1)}_{\mu\nu}$ and $\Sigma^{(0)}_{\mu\nu}$, respectively,  cf.\ \eqref{E:Shear}. 
Another result we obtained is an explicit expression for $\lambda_2$ in terms of the components of the bulk metric, dual to a static configuration of the fluid,  cf.\ \eqref{E:lambda2val}. A similar expression for $\lambda_1$ is
\begin{equation}
\label{E:lambda1val}
	\lambda_1 = \frac{\kappa^{-2}}{4} r_+^{D-3}s(r_+)-\frac{\kappa^{-2}}{8} {\rm Finite}\left[\lim_{r \to \infty}
		\left(\frac{s(r)}{f(r)}\int_{r_+}^{r}\theta_3(x)dx\right) \right],
\end{equation}
where, within the current analysis, $\theta_3$ is an undetermined function.\footnote{
The $\theta_3$ term in \eqref{E:lambda1val} originated in $\partial_{\langle i}\partial_{ j \rangle} \frac{\mu^d}{\epsilon}$ terms which showed up when we took derivatives of Weyl invariant quantities. It was the $\partial_{\langle i}\partial_{j \rangle} \epsilon$ contribution to such second order derivatives which, when evaluated under the constraint equation \eqref{E:epsilonspace}, generated a $\theta_3(r) \partial_{\langle i} \left(u^{\mu} \partial_{\mu} u_{j \rangle}\right)$ term. It seems likely that by an appropriate choice of the $\Sigma^{(n)}_{\mu\nu}$'s with $n > 3$, say, $\Sigma^{(n_0)}_{\mu\nu} = \mathcal{D}_{\langle \mu}\mathcal{D}_{\nu \rangle} \frac{\mu^d}{\epsilon}$ with $\mathcal{D}_{\mu}$ the Weyl covariant derivative of \cite{Loganayagam:2008is}, one should be able to reshuffle the distribution of the $\theta_3$ terms, and get rid of the $\theta_3$ dependence in \eqref{E:Lambdas}. The current formulation of the problem does not allow us to check this possibility explicitly. In the specific case considered in \cite{Erdmenger:2008rm}, $\lambda_1$ was given by \eqref{E:lambda1val} with $\theta_3=0$.}

The choice of basis $\{\Sigma^{(n)}_{\mu\nu} \}$ for the second order contributions to $\langle T_{\mu\nu}\rangle$ is arbitrary and was taken from \cite{Baier:2007ix}. There, $\Sigma^{(0)}_{\mu\nu}$ was chosen to conform with Israel-Stewart theory in the linear regime.
But, one can, for example, work with the alternative basis 
\begin{align}
\begin{split}
\label{E:alternateSigma}
	\widetilde{\Sigma}^{(0)}_{\mu\nu}  &= \Sigma^{(0)}_{\mu\nu} + 2 \Sigma^{(2)}_{\mu\nu}\ , \\
	\widetilde{\Sigma}^{(1)}_{\mu\nu}  &= \Sigma^{(1)}_{\mu\nu} - 4 \Sigma^{(2)}_{\mu\nu}\ , 
\end{split}
\end{align}
which is similar to that of \cite{Bhattacharyya:2008mz}.\footnote{
In \cite{Bhattacharyya:2008mz} the authors considered a CFT with a non flat boundary metric. This allows for an extra contribution to the energy momentum tensor at second order in a derivative expansion, not considered in this work, proportional to the Weyl tensor contracted with the velocity fields. The basis used in equation (1.3)  of \cite{Bhattacharyya:2008mz} and the discussion in \cite{Bhattacharyya:2008mz} suggests an additional relation, similar to \eqref{E:Universal2}, for conformal theories on curved manifolds.
}
Using \eqref{E:Universal2}, we find that in the alternative basis the coefficient of $\Sigma^{(2)}_{\mu\nu}$, $\tilde{\lambda}_2$, vanishes,
\begin{equation}
	\eta \tau_{\pi} \Sigma^{(0)}_{\mu\nu}+\lambda_1 \Sigma^{(1)}_{\mu\nu} + \lambda_2 \Sigma^{(2)}_{\mu\nu}
	=
	\eta \tau_{\pi} \widetilde{\Sigma}^{(0)}_{\mu\nu} + \lambda_1 \widetilde{\Sigma}^{(1)}_{\mu\nu}.
\end{equation}

It is interesting to compare \eqref{E:Universal2} with a weak coupling analysis. Such an analysis has been initiated in \cite{Baier:2007ix} and carried out explicitly for QCD (with 0 and 3 flavors), QED and $\phi^4$ theory in 4 spacetime dimensions and in the absence of a chemical potential in \cite{York:2008rr}.  According to \cite{Baier:2007ix,York:2008rr} the relation \eqref{E:Universal2} does not hold at weak coupling. Instead one finds
\begin{equation}
\label{E:Universalweak}
	\lambda_2 = - 2 \eta \tau_{\pi}.
\end{equation}
In the alternative basis, \eqref{E:Universal2} and \eqref{E:Universalweak} can be translated into the statement that $\tilde{\lambda}_2$ increases from $-4(\eta \tau_{\pi}-\lambda_1)$ at weak coupling to 0 at strong coupling (and infinite $N$). The perturbative computation of \cite{York:2008rr} for QCD indicates that, in the alternative basis, $\tilde{\lambda}_2$ increases with the coupling, in agreement with our previous statement.

To obtain \eqref{E:Universal2}, we had to assume that the theory at hand is conformal, that its bulk dual can be truncated to an action of the form \eqref{E:action}, that the $U(1)$ symmetries are not spontaneously broken, and that the various fields have an asymptotic expansion as in \eqref{E:fsAasymptotics}. The main reason why we worked with the asymptotic expansion in \eqref{E:fsAasymptotics} is that, otherwise, it would be difficult to handle the extra terms which arise when holographically renormalizing the theory. 

Some of these restrictions could be easily removed. For example, we may allow for sources for the scalars which correspond to operators with dimension smaller than four. In such case, the main change in our computation is that the energy-momentum tensor may be sourced by gradients of the sources
(as was the case for the dilaton in \cite{Bhattacharyya:2008ji}). This would simply mean that the right hand side of \eqref{E:epsilonspace} and \eqref{E:epsilontime} needs to be modified by adding to it appropriate derivatives of the scalars. Since the merit of \eqref{E:epsilonspace} and \eqref{E:epsilontime} is in converting space-time derivatives of the energy density into spacetime derivatives of the velocity field, most of our analysis will remain unchanged. 

Perhaps a bigger challenge to \eqref{E:Universal2} is a non-conformal theory. Recall that we relied heavily on conformal invariance to obtain the form of the first order vector expressions in \eqref{E:O1vecs} and to set the $\overline{\Lambda}_n$'s to zero in \eqref{E:EOM2ndshortest}. If these coefficients were not zero, we would not have been able to obtain \eqref{E:Lambda}. Curiously, even if the $\overline{\Lambda}_n$'s did not vanish, we could still have written
\begin{equation}
\label{E:NonCLambda}
	4 \Lambda_1+ \Lambda_2 - 2 \Lambda_0 - \overline{\Lambda}_2
	=
	 4 \left(\left(r^{D-2}-r_+^{D-2}\right)F(r)\right)^{\prime}\ ,
\end{equation}
which suggests a relation of the form
\begin{equation}
\label{E:UniversalNC}
	4 \lambda_1 + \lambda_2 - 2 \eta \tau_{\pi} - \overline{\lambda}_2 = 0\ ,
\end{equation}
where the $\overline{\lambda}_i$ are the second order transport coefficients associated with $\overline{\Sigma}^{(i)}_{\mu\nu}$ when expanding a non conformal stress energy tensor in a hydrodynamic expansion,
\begin{equation}
	\langle T_{\mu\nu} \rangle
	=
	\epsilon u_{\mu}u_{\nu} + p P_{\mu\nu}
	-\eta \sigma_{\mu\nu} - \zeta P_{\mu\nu} \partial_{\lambda}u^{\lambda}
	+\eta\tau_{\pi} \Sigma^{(0)}_{\mu\nu}
	+\lambda_i \Sigma^{(i)}_{\mu\nu} + \overline{\lambda}_i \overline{\Sigma}^{(i)}_{\mu\nu}+\ldots\ .
\end{equation}
Of course, once the system is non-conformal there will be further corrections to \eqref{E:EOM2nd} coming from the bulk viscosity $\zeta$ and pressure $p$. It remains to be seen if there is a relation of the form \eqref{E:Universal2}  or \eqref{E:UniversalNC} that holds for non-conformal theories. Perhaps a good starting point for such an exploration are the relatively simple non-conformal backgrounds constructed in \cite{Gubser:2008yx}.

\section*{Acknowledgements}
We thank N.~Banerjee, J.~Bhattacharya, S.~Bhattacharyya, G.~Cardoso, S.~Dutta, J.~Erdmenger, M.~Ka\-min\-ski, A.~Karch, R.~Loganayagam, S.~Minwalla, T.~Mohaupt, I.~Sachs, D.~Son and P.~Sur\'owka for discussions and correspondence.
M.~H.~is supported in part by the European Community's Human potential program under contract MRTN-CT-2004-005104 ``Constituents, fundamental forces and symmetries of the universe'', the Excellence Cluster ``The Origin and the Structure of the Universe'' in Munich and the German Reseacrh Foundation (DFG) within the Emmy-Noether-Program (grant number: HA 3448/3-1). A.~Y. is supported by the Department of Energy under Grant No. DE-FG02-91ER40671.

\begin{appendix}
\section{Holographic renormalization}
\label{A:HR}
By varying the bulk action $S_{\rm total}$ with respect to the boundary theory metric $g^{(0)}_{\mu\nu}$, one obtains the stress energy tensor of the boundary theory
\begin{equation}
	\langle T_{\mu\nu} \rangle = \frac{2}{\sqrt{-g^{(0)}}}\frac{\delta S_{\rm total}}{\delta g^{(0)\,\mu\nu}}.
\end{equation}
The action $S_{\rm total}$ is composed of the bulk action \eqref{E:action}, evaluated on-shell, a Gibbons-Hawking boundary term, and a counterterm action 
\begin{equation}
	S_{\rm ct} = \int_{r=R_0} d^dx \mathcal{L}_{\rm ct}\ ,
\end{equation}
which renders $S_{\rm total}$ finite when evaluated on-shell. The contribution of the Einstein part of the bulk action and the Gibbons-Hawking term to $\langle T_{\mu\nu} \rangle$ together with a boundary cosmological constant was given explicitly in \eqref{E:BtB}.

Let us rewrite the counterterm action in the form
\begin{equation}
	S_{\rm ct} =  \int  d^{d}x \sqrt{\gamma} \left(\mathcal{L}_0 + \mathcal{L}^{\mu_1}A_{\mu_1}
		+\mathcal{L}^{\mu_1 \mu_2} A_{\mu_1} A_{\mu_2}\ ,
		+\ldots \right)\, ,
\end{equation}
where the dependence of the Lagrangian on the gauge field has been made explicit. The $\mathcal{L}_{\mu_1 \ldots \mu_n}$'s are constructed from the various scalars and derivatives and can be thought of as acting on the terms on their right. We have denoted the metric on a slice of constant $r=R_0$ by $\gamma_{\mu\nu}$. Near the boundary $\gamma^{\mu\nu} \sim R_0^{-2}$, $\sqrt{\gamma} \sim R_0^{d}$. From \eqref{E:fsAasymptotics}, $A^a_{\mu} \sim R_0^{-d+2}$ and assuming that we are not turning on relevant operators in the conformal boundary theory, $\phi^\alpha \sim R_0^{m_{\alpha}}$ with $m_{\alpha} \leq 0$.

Suppose $\mathcal{L}_{\mu_1 \ldots \mu_n} \sim R_0^m$. If the $\mathcal{L}_{\mu_1 \ldots \mu_n} $ term is to have any contribution to the counterterm action in the $R_0\to\infty$ limit, we need that $m-d(n-1) \geq 0$. On the other hand, 
$\mathcal{L}_{\mu_1 \ldots \mu_n}$ is composed of the scalars and their derivatives. By our previous discussion this means that $m \leq 0$. Thus, only the $\mathcal{L}_0$ and $\mathcal{L}_{\mu}$ terms can contribute to the  non-vanishing parts of $S_{\rm ct}$ in the $R_0 \to \infty$ limit. 
Gauge invariance and the absence of charged scalars imply that the $\mathcal{L}^{\mu}$ term involves a derivative acting on $A_{\mu}$. In this case $\mathcal{L}^{\mu}$ involves three derivatives, so that $\mathcal{L}^{\mu}A_{\mu} \sim R_0^{-d-3}$. Hence, this term also vanishes in the $R_0 \to \infty$ limit.
We conclude that the contribution of the counterterm action to the boundary theory energy momentum tensor will not include the various gauge fields.

\end{appendix}

\bibliographystyle{JHEP}
\bibliography{etaovers}

\end{document}